# Assessment of tear film using videokeratoscopy based on fractal dimension

Clara Llorens-Quintana[1], MSc, and D. Robert Iskander[1], PhD, DSc

[1]Department of Biomedical Engineering, Wroclaw University of Science and Technology,

Wroclaw, Poland

Corresponding author:

Clara Llorens-Quintana

Department of Biomedical Engineering,

Wroclaw University of Science and Technology

Wybrzeze Wyspianskiego 27, 50-370 Wroclaw, Poland

Phone: +48 71 320 4661

E-mail: clara.llorens.quintana@pwr.edu.pl

Financial interest of the authors: none

# Abstract

Significance: The proposed automated approach for estimating the quality of the tear film closes the gap between the manual and automated assessment, translating the high speed videokeratoscopy technology from scientific laboratories to a clinical practice.

Purpose: To develop and test a new method for characterizing tear film surface quality with high speed videokeratoscopy utilizing a fractal dimension approach.

Methods: The regularity of the reflected pattern in high speed videokeratoscopy (E300, Medmont) depends on tear film stability. Thus determining tear film stability can be addressed by estimating the fractal dimension of the reflected pattern. The method is tested on 39 normal subjects. The results of the fractal dimension approach are compared with those obtained using previously proposed automated method, based on a gray level co-occurrence matrix approach, and with subjective results obtained by two operators which were assessing the video recordings in ideal conditions.

Results: fractal dimension method was less affected by eye movements and changes in the videokeratoscopic image background than gray level co-occurrence matrix method. Median difference of the non-invasive break-up time between manual and automated methods was 0.03 s (IQR = 4.47 s.) and 0.0 s (IQR = 2.22 s.) for gray level co-occurrence matrix and fractal dimension approaches, respectively. Correlation coefficient with manual non-invasive break-up time was $r2 = 0.86$ ($p < 0.001$) for gray level co-occurrence matrix approach, and $r2 = 0.82$ ($p < 0.001$) for fractal dimension approach. Significant statistical difference was found between non-invasive break-up measurements of manual and gray level co-occurrence matrix method ($p = 0.008$).

Conclusions: The proposed method has the potential to characterize tear film dynamics in more detail compared to previous methods based on high speed videokeratoscopy. It showed good correlation with manual assessment of tear film.

**Introduction**

The tear film is a thin, but complex layer of fluid that covers the cornea and conjunctiva. It has an important role in maintaining the health of the eye, being responsible for supplying nutrients, removing debris from the cornea, protecting the eye from infections, preventing the desiccation of the corneal surface and providing a smooth refractive surface for a proper image formation.[1,2] As a dynamic fluid it undergoes reformation with each blink. The reformation phase (sometimes referred to a build-up phase[3]) is followed by a period of stability and, if the eye is opened for a sufficiently long time or tears are not sufficiently stable, deformation (or break-up) will occur prior to the next blink. A quantitative or qualitative deficiency of tear film could lead to dry eye disease.[4] Thus, assessing tear film quality is important in a clinical practice.

One approach to determine tear film quality is based on evaluating its stability. The main measure of tear film stability is the tear film break-up time that is defined as the time from the last blink until the first appearance of some instability in tear characteristics.[5] Non-invasive measurement of break-up time was recommended as the best method for assessing tear film stability for clinicians.[6] The preferred technique should be objective, with high specificity and reproducibility. Different techniques have been used to assess tear film stability such as confocal microscopy, aberrometry, TearScope white light interferometry lighting system, lateral shearing interferometry or videokeratoscopy.[7,8,9] The videokeratoscope is a corneal topographer based on pattern projection principle where a set of concentric rings (Placido disk pattern) is projected onto the corneal surface. The topography of the cornea is reconstructed based on the size, shape and regularity of the reflected image. As the outer layer of the eye is the tear film, the Placido disk pattern reflection will depend on tear film surface stability.

Placido disk high speed videokeratoscopy,[10] which extends the traditional static videokeratoscopy to a dynamic capture, has been proposed as a potential clinical tool to assess Tear Film Surface Quality. It is easy to use, clinicians use it on the daily clinical practice and it is easily accessible.[11] In addition, it covers a large surface of the eye (up to 10 mm in diameter) and it is not as affected as others techniques by eye movements.[12] Some approaches have been proposed to characterize the pre-corneal and pre-lens tear film using high speed videokeratoscopy (see Table 1). First studies that used high speed videokeratoscopy derived the dynamics of tear film from the reconstructed data of corneal topography (e.g., using regularity and asymmetry indices and corneal power). However, several shortcomings have been reported in such tear film characterization[13,14] because an estimator of a topographic feature requires good quality of tear film in the first place. Consequently, image processing techniques that assess tear film dynamics from the raw videokeratoscopic images were proposed. These methods were able to describe the tear film dynamics in different groups of subjects including those with healthy tear film, diagnosed with dry eye disease (aqueous deficiency or Meibomian gland dysfunction) and contact lens wearers.[15,16,17] Nowadays there are commonly available Placido disk high speed videokeratoscopy instruments that include an option for automated tear film analysis. However, there is still a lack of correlation between the proposed algorithms and the subjective, manually operated, evaluation of the recorded images of high speed videokeratoscopy, such as in the case of Oculus Keratograph[18] and Medmont E300.[19] It is worth noting that in the case of Placido disk high speed videokeratoscopy, the acquired data correspond to a given physical state of tear film lipid layer on the eye, where instabilities in tear film directly relate to disturbances of the recorded Placido disk pattern. The aim of both manual operator-based and automated schemes is to unambiguously determine places and times at which those disturbances occur. That is why correspondence between subjective and objective measures derived from high speed videokeratoscopy images is crucial.

Hence, the need of finding a new approach for analyzing the reflected pattern that correlates with the subjective assessment arises. The basis of the problem lies on understanding when a clinician reports that the tear film is not stable. Clearly, when the reflected Placido disk pattern is regular, tear film is considered to be smooth and stable. In Figure 1, different situations of tear film destabilization are shown. When the tear film becomes unstable the reflected rings are uneven, washed out and even missing. Therefore, the problem of determining tear film stability can be addressed from the point of view of estimating the texture of the reflected pattern. This can be done by means of computing the fractal dimension of the reflected rings inasmuch as fractal dimension has been shown to correlate with human perception of texture and describes line irregularity.[20] Thus, fractal dimension has a direct correspondence with local morphological changes of the reflected pattern unlike the previous estimates that were describing tear film with global parameters that are unable to directly characterize the features observed in the distorted ring pattern.

The aim of this work was to develop and clinically evaluate a new algorithm based on a fractal dimension approach to characterize tear film dynamics described by the morphological changes of high speed videokeratoscopy pattern reflections in healthy subjects, to equip clinicians with a technique that they could directly relate to and agree with.

Table 1
A chronological account of HSV tear film developments

| Authors (year) | Sampling frequency | Modality | TF quality indicator | Studied phase of TF |
|---|---|---|---|---|
| Nemeth et al. (2002)[3] | 4 Hz | Topography | SRI, SAI and corneal power | Build up |
| Goto et al. (2003)[21] | 1 Hz | Topography | Corneal power | 10 s after blink |
| Montés-Micó et al. (2004)[22] | 1 Hz | Topography | Corneal aberrations | 15 s after blink |
| Kojima et al. (2004)[23] | 1 Hz | Topography | SRI and SAI | 10 s after blink |
| Iskander et al. (2005)[13] | 50 Hz | Topography/Image analysis | RMS/variance of ring counts | Build up/break up |
| Zhu et al. (2006)[24] | 50 Hz | Topography | Surface difference height maps | IBI (NBC) |
| Kopf et al. (2008)[25] | 25 Hz | Image analysis | Variance of ring counts | 8 s after blink |
| Alonso-Caneiro et al. (2009)[26] | 25 Hz | Image analysis | Coherence of reflected rings | 8 s after blink |
| Alonso-Caneiro et al. (2011)[27] | 25 Hz | Image analysis | Pattern regularity-block featured | 30 s SBC NBC |
| Gumus et al. (2011)[28] | 1 Hz | Image analysis | Brightness of reflected rings | 6 s after blink |
| Alonso-Caneiro et al. (2013)[29] | 25 Hz | Image analysis | Texture analysis-GLCM | 30 s SBC |
| Downie (2015)[19] | 4 Hz | Image analysis | Proprietary (Medmont) | 23 s SBC |
| Szczesna-Iskander et al. (2016)[30] | 30 Hz | Image analysis | Texture analysis-GLCM (homogeneity) | 50 sec SBC |

SRI: Surface regularity index
SAI: Surface asymmetry index
GLCM: Gray level co-occurrence matrix
SBC: Suppressed blinking conditions. The patient blinks and holds the eyes open as much as possible
NBC: Natural blinking conditions

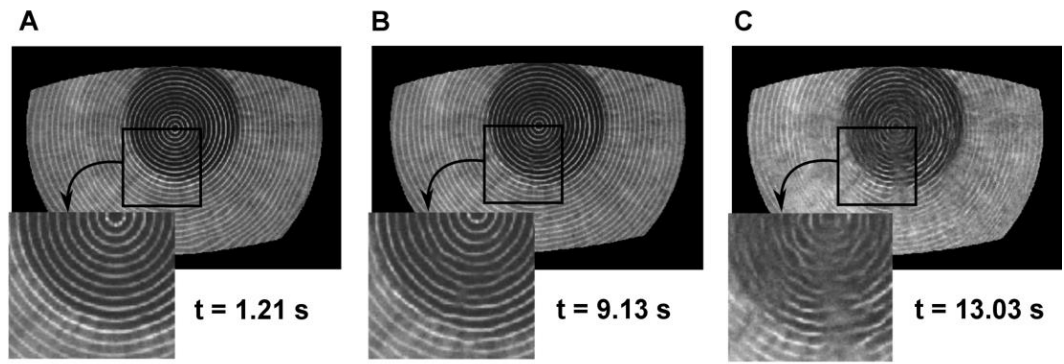

**Figure 1.** The appearance of Placido disk reflections for different stages of tear film dynamics. Stable tear film (A, $t$ = 1.21 s post-blink), tear film destability (B, $t$ = 9.13 s post-blink), and tear film deterioration (C, $t$ = 13.03 s post-blink). A mask corresponding to an average Caucasian anterior eye shape is applied (see Methods section).

**Methods**

*Background*

The fractal dimension is a measure of texture roughness that in a digital image depends on its pixel organization.[32] It has become popular in the field of image analysis since it correlates quite well with the human perception of texture and is relatively insensitive to changes in image intensity and scaling.[31] There are several alternative definitions of fractal dimension. Similarly, there are several approaches to estimate fractal dimension of a digital image.[33,34] The box-counting method is one of the most commonly used due to its easy implementation and simplicity of the computation. Also, the box-counting method has shown to provide good approximations of theoretical fractal dimension.[35] For its computation, an image is covered by a grid of boxes of size $r$ and the number of boxes that contain a portion of the object $N(r)$ are counted; this procedure is repeated decreasing the size of the boxes successively until no changes are observed. Fractal dimension is given by the relation between $N(r)$ and $r$ as:

$$FD = -\lim_{r \to 0} \frac{\log(N(r))}{\log(r)}$$

From the $N(r)$ vs. $r$ log-log plot we obtain a curve with slope that provides an estimate of fractal dimension.

In order to relate the $FD$ measure with the stability of the tear film, first we transform the videokeratoscopic image from Cartesian to polar coordinates. In this way the Placido disk pattern is represented by a set of semi-horizontal patterns. For a straight line, $FD$ equals to one (the same as its topological dimension) and it increases as it becomes irregular, ranging between one and two. In the case of an incomplete line $FD$ is lower than its topological dimension, ranging between zero (no line) and one (see Figure 2). The regularity of the reflected Placido disk pattern in high speed videokeratoscopy, depends on tear film stability. Hence, a disruption or destabilization of tear film will produce an irregular reflected Placido disk pattern with either uneven or faded rings. To determine when the tear film is stable an empirically evaluated value of $k$ = 0.18 was set, so that if $FD$ value of the reflected rings satisfy: $1 + k > FD > 1 - k$, the tear film is considered stable.

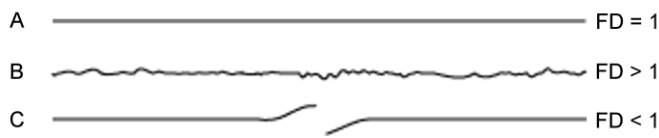

**Figure 2.** The concept of fractal dimension used in the context of high speed videokeratoscopy. A—straight line, $FD = 1$, equivalent to topological dimension; B — distorted line, $FD > 1$, larger than the topological dimension; C — broken line, $FD < 1$, smaller than the topological dimension.

*Image processing*

For this work, E300 videokeratoscope (Medmont Pty., Ltd, Melbourne, Australia) was used with a sampling frequency of 13 frames per second. The dynamic videokeratoscopy acquisition is composed of a set of gray-scale intensity images with size $648 \times 572$ pixels. Before analyzing the images, the first step is to detect those frames affected by blinks or large eye movements, which do not contain useful information, and remove them from further analysis (Figure 3a). The remaining consecutive frames will form the sequence where the analysis will be performed, called inter blink interval.[26] Once the inter blink interval is identified, the area of analysis is defined as a corneal surface corresponding to an average Caucasian eye with a diameter of 10 mm and centered at the instrument axis (Figure 3b).[36]

In some frames the central axis of the videokeratoscope does not correspond to the geometric center of the concentric reflected rings. To overcome this limitation, the Hough circular transform, which is an algorithm to detect circular features in a digital image, is used to identify the inner circle of the reflected pattern and its center is taken as the center for the future analysis (Figure 3c). The effect of upgrading the center on the fractal dimension based Tear Film Surface Quality can be seen in Figure 4.

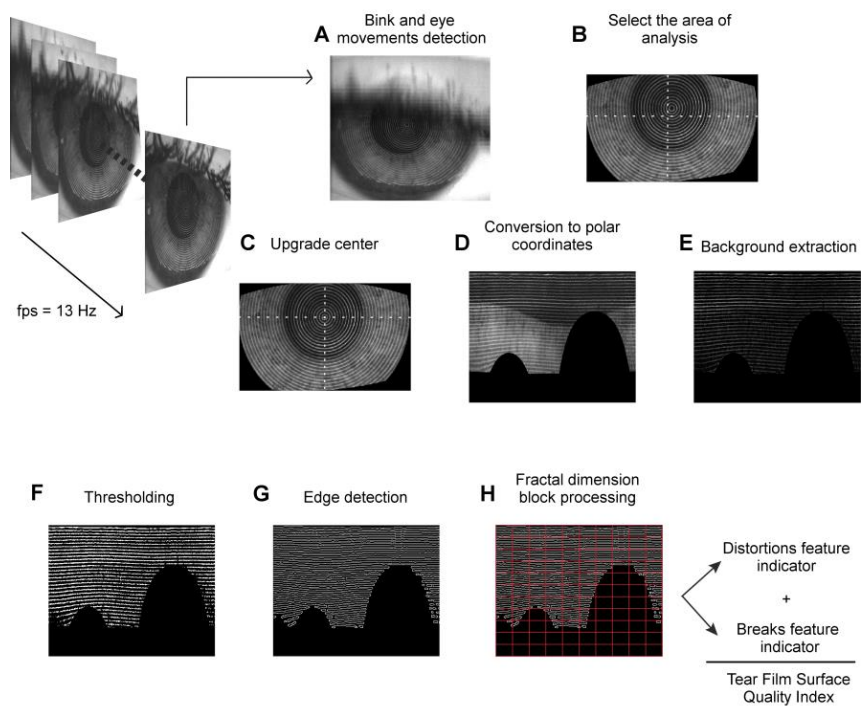

**Figure 3.** The schematics of the algorithm used for estimating the fractal dimension parameters in HSV. The size of the grid in the fractal dimension block processing step H is given for illustrative purpose and it does not correspond to the actual size of the block (i.e., 9 × 10 pixels).

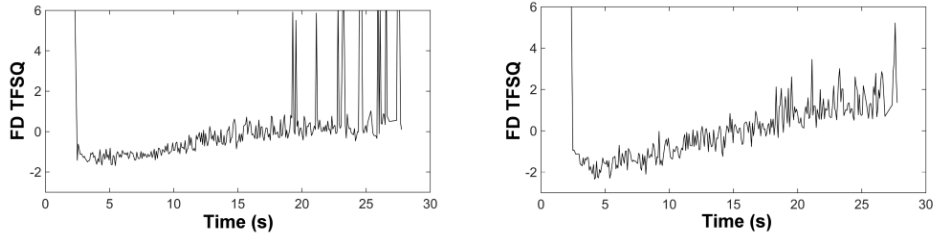

**Figure 4.** The effect of image centration on the resulting estimator of TFSQ. Image sequence centered at the instrument axis (left) and centered at the geometrical center of the inner Placido disk ring (right).

In order to study the morphology of the reflected rings we need to isolate them from the eye background (iris and pupil) and demarcate their edges. First, the image is transformed from Cartesian to polar coordinates (Figure 3d), suppressing the influence of the Placido disk pattern orientation and converting the concentric circles into horizontal lines. Then, the background of the polar coordinate image is subtracted by creating a background approximation using morphological operations (image opening, using a 3 pixels diameter disk as a structural element) and subtracting it from the original polar image (Figure 3e). In order to enhance the morphology of the reflected rings a global thresholding is performed using standard Otsu's method with two classes.[37] Otsu's algorithm assumes that the image contains information related to two classes and finds the optimal threshold to separate both so that their interclass variability is minimal. After thresholding, the corresponding binary image is obtained (Figure 3f). On the resulting image, the Canny edge detection algorithm[38] is performed (Figure 3g). Canny algorithm is the first derivative of a Gaussian that approximates the operator that optimizes the product of signal-to-noise ratio and localization.

Once the edges of the reflected Placido disk pattern are isolated, the transformed polar image is divided into non-overlapping blocks of equal size of $10 \times 9$ pixels, determined empirically for the E300 instrument. The size of the blocks is chosen based on the separation between the reflected rings, so that they contain sufficient amount of information with enough spatial resolution. The box-counting method is used to compute $FD$ in each of the blocks within the area of analysis (Figure 3h). From this analysis two features and one index are extracted from each frame (Figure 3i), forming time series sequences:

a) Distortion Feature Indicator: in the case of smooth and regular tear film the edges of the reflected rings of the polar coordinate image should approach a set of horizontal lines. Irregularities of tear film will cause uneven lines. Blocks containing uneven lines will have higher $FD$ than those with straight lines. For $N_{DFI}$ blocks with $FD$ values above $1 + k$ (above the defined stable interval for the tear film, see background section), the Distortion Feature Indicator is computed as:

$$DFI = \sum_{N_{DFI}} \frac{1}{2 - FD(N_{DFI})}$$

b) Breaks Feature Indicator: destabilization of tear film can also lead to a break of the rings, so that the edges of the polar image will be discontinued (see Fig. 2). $FD$ of blocks with discontinued lines (or even no lines) will be lower than those with complete lines. For $N_{BFI}$ blocks with $FD$ values below $1 - k$ (below the defined stable interval for the tear film, see background section) the Breaks Feature Indicator is computed as:

$$BFI = \sum_{N_{BFI}} \frac{1}{FD(N_{BFI})}$$

c) Tear Film Surface Quality index: in order to describe the quality of the tear film, it is desirable to consider the effect of both features (distortions and breaks), so that fractal dimension based Tear Film Surface Quality index is obtained from the addition of distortion and breaks features indicators.

In other words, the breaks are characterized with $FD$ values ranging from zero to $1 - k = 0.82$ while distortions from $1 + k = 1.18$ to two. The tear film quality indicators $DFI$ and $BFI$ are summations of the reciprocals of $FD$ estimates for the breaks and distortions intervals, respectively. To ensure the equivalence of $DFI$ with respect to $BFI$, the $FD$ in that interval is subtracted from two.

Finally, both features and the resulting Tear Film Surface Quality index are statistically normalized by first subtracting the mean value and later by dividing them by the respective standard deviations. Then, lower values of $DFI$, $BFI$ and Tear Film Surface Quality correspond to a good quality tear film, whereas higher values correspond to a worse quality tear film.

Although the chosen area of analysis takes into account the possible shadows produced by the eye lashes, in some cases, where a subject has long eye lashes or narrow palpebral aperture, shadows can fall inside the area of analysis leading to a misrepresentation of breaks feature. To overcome this limitation, blocks linked to the superior boundary of the area of analysis, with $FD < 0.82$ are not taken into account for the analysis.

A representative example of $DFI$, $BFI$ and Tear Film Surface Quality index time series along the inter blink interval and their correspondence to the recorded raw images and $FD$ block values (in the transformed polar image), are shown in Figure 5. Time $t_1$ corresponds to the tear film build-up phase; at this point the percentage of breaks is greater than the percentage of distortions, this could be related to the lipid layer moving upwards. In $t_2$ the tear film is already stable, with a lower percentage of breaks whereas distortions remain constant. Time $t_3$ corresponds to the estimated break-up time caused by an increase of distortions in the Placido disk pattern.

*Subjects and data acquisition*

In order to test the performance of the proposed algorithm, data from 39 Caucasian subjects was acquired with E300 videokeratoscope, all of them were no contact lens wearers. There were 24 female and 15 male subjects aged between 20 and 40 years. The study followed the tenets of Declaration of Helsinki and all participants signed an informed written consent. The study was approved by the Ethic Committee of the Wroclaw Medical University. Eye health was assessed by an experienced clinician. Participants were included if there was no evidence or history of ocular tissue anomaly, ocular surgery, ocular infection or inflammation, dry eye, allergy, or any ocular surface or systemic disease that may affect the tear film.

The clinical assessment of the tear film was performed on two different days. On the first day, dry eye was evaluated with a standard clinical assessment of signs and symptoms that included (in order of performing): Ocular Surface Disease Index questionnaire, grading of corneal and conjunctival staining using fluorescein and lissamine green, respectively, grading of meibomian gland dysfunction performed via slit lamp examination and fluorescein break-up time. Dry eye was diagnosed if at least one symptom (Ocular Surface Disease Index questionnaire score higher than 15) and one sign (fluorescein break-up time lower than 7 seconds, corneal or conjunctival staining score greater than 2 and meibomian gland dysfunction score greater than 2 according to the Efron's grading scale) were present. On the second day both eyes of each subject were examined twice (alternating eyes) with the E300 videokeratoscope, with a break of three minutes between each measurement.

In case of substantial discrepancies (visible during recordings) a third measurement would be performed. However, that was not necessary for any of the subjects considered in the study. Participants were asked to fixate the central target of the instrument, blink a few times and maintain their eyes open as long as possible (suppress blinking conditions), for a maximum time of 40 seconds.

The acquired recordings were then evaluated with the proposed automated method as well as with a previously proposed method based on textural analysis using a gray level co-occurrence matrix approach.[29] To estimate the break-up time with both automated methods (fractal dimension and gray level co-occurrence matrix), the Tear Film Surface Quality time series sequences were fitted with a constrained bilinear model and the point of constraining was taken as the estimated break-up time. Further manual non-invasive break-up time was estimated by two operators, independently, examining the morphological changes in Placido disk pattern reflections in the video sequence using a custom written graphic user interface that provides the means to visualize and review the recording and to annotate the place and time of a disturbance in the Placido disk pattern that is considered to be linked with tear film break-up. A protocol for the manual non-invasive break-up time estimation was carefully followed by both operators: performing an initial review of the video and observing the blinking patterns, identifying the inter blink interval of interest and identifying the non-invasive break-up time frame as the first significant or noticeable disturbance in the rings pattern that has an evolution in time.[26] Note that this is an ideal scenario, where both operators had unlimited time to evaluate the recordings, that may not be the case in the clinical practice where clinicians only can see the recorded video once at full-speed.

For the analysis of the data, repeated measurements of the same eye were averaged and only measurements of right eyes were considered. The average of manual non-invasive break-up time of both observers was taken as a unique manual non-invasive break-up time. The statistical analysis included standard descriptive measures, normality test (Shapiro–Wilk), non-parametric hypothesis testing (Friedman test) and correlation analysis (Spearman's correlation coefficient).

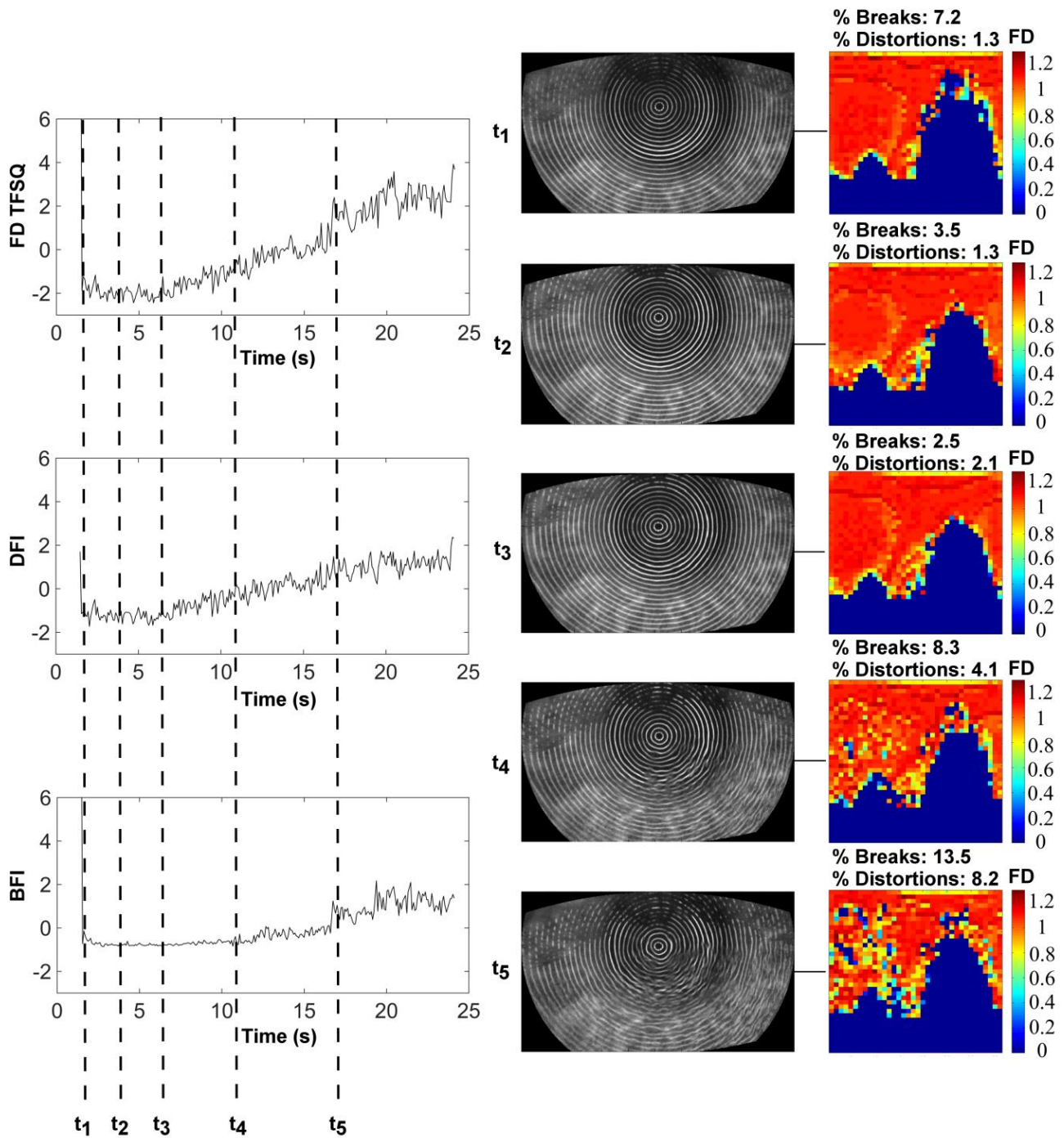

**Figure 5.** Example of $DFI$, $BFI$ and TFSQ time series (left column) and corresponding Placido disk images (center column) acquired at the time instances indicated on the time series plots with dashed vertical lines $(t_1, t_2, ..., t_5)$. Right column shows the corresponding $FD$ estimates in the polar coordinate systems.

**Results**

Participants' demographics together with the diagnostic scores are shown in Table 2.

Table 2

Study participants demographics and the diagnostic scores ($n = 39$). SD denotes standard deviation and MAD the median average distance

| Characteristic | Group Mean or Median*, (range) | Group SD or MAD* |
|---|---|---|
| Age [years] | 25 (21-39) | 3.4 |
| OSDI score | 8.1 (0-25) | 7.3 |
| Corneal staining score | 0.0*(0-1) | 0.4* |
| Conjunctival staining score | 1.0* (0-2) | 0.5* |
| Meibomian gland dysfunction score | 0.0* (0-2) | 0.4* |
| FBUT [s] | 12.6 (7-26.5) | 4.2 |

Figure 6 shows some representative examples of fractal dimension based Tear Film Surface Quality index along the inter blink interval. For Figure 6a, b, and c the change in the slope of the data determines the point where the tear film starts to deteriorate (corresponding to the estimated non-invasive break-up time) showing a typical dynamic behavior of the tear film with a short build-up, followed by a stable phase that ends when the tear film gets worse. A different situation can be observed in Figure 6d where, although a build-up phase might be present, there is no break-up point, meaning that the tear film is more or less stable along the entire inter blink interval.

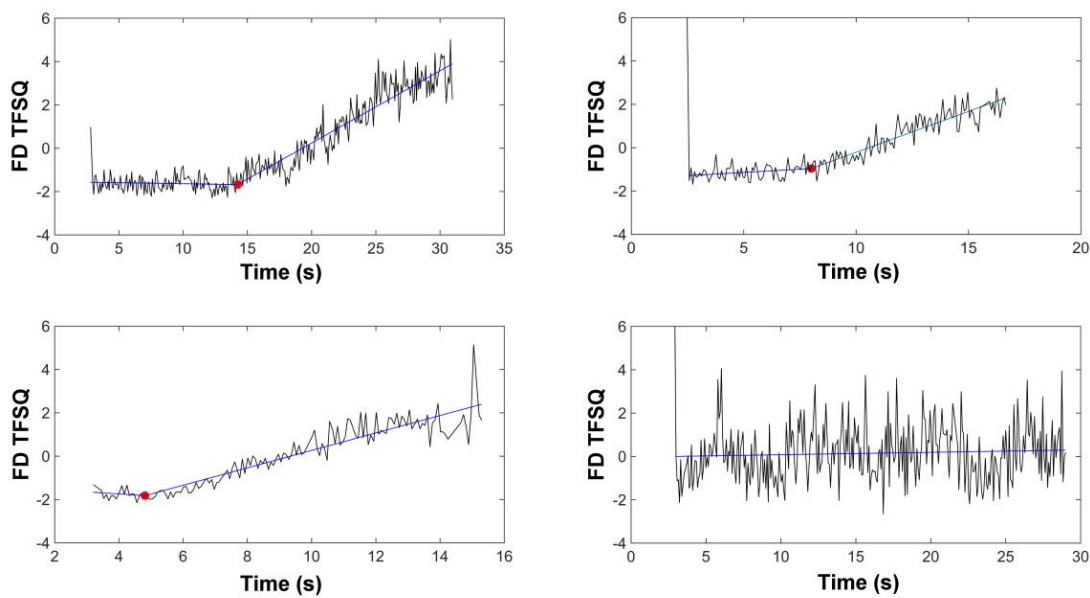

**Figure 6.** Examples of fractal dimension based TFSQ (FD TFSQ) estimators showing different behavior of tear film for different inter blink intervals together with linear or bilinear best fits (blue lines). Red dots indicate the estimated tear film break-up time. Refer to text for further details on the interpretation of the results.

During a dynamic data acquisition micro-movements of the eye are frequent. This could be an issue when characterizing the tear film dynamics. An example of this situation is shown in Figure 7 for fractal dimension and gray level co-occurrence matrix approaches. It can be seen that the fractal dimension based Tear Film Surface Quality estimator is less affected by eye movements than that based on gray level co-occurrence matrix approach, where eye movements lead to impulsive (high peaks) changes in Tear Film Surface Quality.

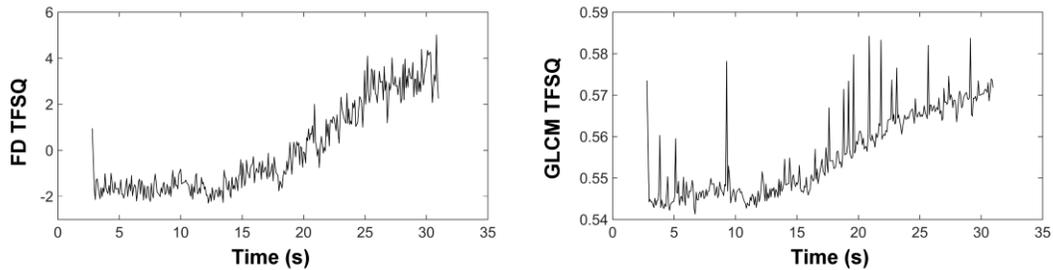

**Figure 7.** The effect of eye movements on the fractal dimension based TFSQ estimator (left, FD TFSQ) and on that based on gray level co-occurrence matrix (right, GLCM TFSQ).

When measuring Tear Film Surface Quality with a videokeratoscope that operates with a Placido disk in form of a small cone, the pupil size of the subject usually constricts due to the proximity of the highly illuminated Placido disks to the eye. There are also pupil variations related to accommodation fluctuations. Those changes in the acquired image background could lead to inaccurate results of tear film dynamics. An example of this situation for fractal dimension and gray level co-occurrence matrix approaches is shown in Figure 8, where the fractal dimension based Tear Film Surface Quality estimator shows to be less affected by changes in the pupil size than that based on gray level co-occurrence matrix.

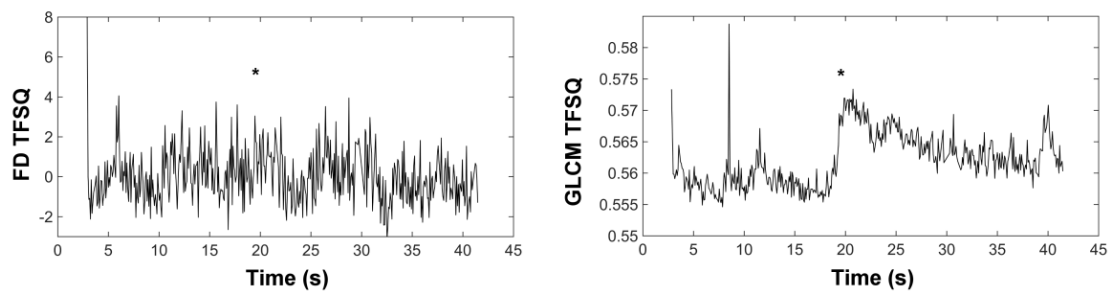

**Figure 8.** The effect of a substantial pupil size changes (indicated by an asterisk) on the fractal dimension based TFSQ estimator (left, FD TFSQ) and on that based on gray level co-occurrence matrix (right, GLCM TFSQ).

Differences between estimated manual non-invasive break-up time and automated non-invasive break-up time were not normally distributed as assessed by Shapiro–Wilk test ($p = 0.009$ for gray level co-occurrence matrix vs. manual and $p < 0.001$ for fractal dimension vs. manual). Median difference between the manual non-invasive break-up time and that estimated automatically was 0.03 s for the gray level co-occurrence matrix approach and 0.00 s for that based on fractal dimension approach. The interquartile range was 4.47 s and 2.22 s for the gray level co-occurrence matrix and fractal dimension based estimators, respectively (see Figure 9). Friedman test revealed a significant difference between the non-invasive break-up time assessed by the three methods: $\chi^2 = 8.389$, $p = 0.015$. Post hoc analysis with Wilcoxon signed-rank tests with a Bonferroni correction showed significant difference between the groups: manual non-invasive break-up time and automated gray level co-occurrence matrix non-invasive break-up time ($Z = 3.099$, $p = 0.008$) and automated gray level co-occurrence matrix non-invasive break-up time and automated fractal dimension non-invasive break-up time ($Z = 3.099$, $p = 0.002$).

Spearman's correlation coefficient between manual non-invasive break-up time and that estimated with the gray level co-occurrence matrix approach was $r^2 = 0.86$ ($p < 0.001$), and $r^2 = 0.82$ ($p < 0.001$) between manual non-invasive break-up time and that estimated with the fractal dimension approach (see Figure 10).

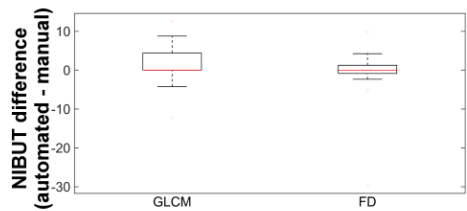

**Figure 9**. Box plot of NIBUT difference between manual and two considered automated methods (based on gray-level co-occurrence matrix (GLCM) and the fractal dimension (FD)).

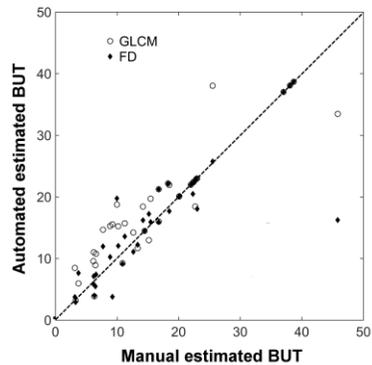

Figure 10. Correlation between manually estimated NIBUT and automatically estimated NIBUT with the gray-level co-occurrence matrix (GLCM) approach (circles) and the fractal dimension (FD) approach (diamonds). The solid line indicates the unity between the two measures.

Tear film dynamics has intrinsically wide variability as it depends on all cycles of tear film formation, particularly in the regime of suppressed blinking. Subjective assessment of high speed videokeratoscopic recordings entails an additional inter-observer variability that will have an impact on the repeatability of the measurements. The group averaged within-observer coefficient of variation between repeated subjective non-invasive break-up time measurements (each video recording seen twice) was 12.79%. By using an automated algorithm this variability is avoided.

**Discussion**

We have proposed and evaluated a new methodology to estimate the quality and dynamics of the tear film from high speed videokeratoscopy measurements. In the proposed method, Tear Film Surface Quality index is derived from a textural analysis of the reflected Placido disk pattern, based on the calculation of its fractal dimension. From this analysis two features ($BFI$ and $DFI$) and one index (Tear Film Surface Quality index) are extracted. $BFI$ is related to discontinuities or fading of the reflected pattern whereas $DFI$ is related to an uneven and irregular reflected pattern. Therefore, each of the features relates to a different type of tear film destabilization phenomena. It is desirable to consider both features when computing Tear Film Surface Quality index, for now both features have the same weight for Tear Film Surface Quality estimation.

In the traditional way of assessing tear film stability (i.e., the fluorescein break-up time) a drop of fluorescein is instilled into the eye. The break-up time is the time until the first black spot appears. However, it is likely that before the appearance of black spots, the surface of tear film could be already destabilized. Different studies[39,40] have reported that the disruption of the tear film may be governed by different mechanisms, all of them contributing to the tear film break-up time. This could explain the difference between $BFI$ and $DFI$ behaviors and how they describe destabilization of the tear film.

The fractal dimension based Tear Film Surface Quality index has shown to be sensitive enough to identify up to three different phases of tear film dynamics: the build-up, which is the time from the blink until the tear film achieves its optimum quality, a stability phase and the break-up. Fractal dimension based Tear Film Surface Quality is less affected by changes in the background of high speed videokeratoscopy images (e.g. changes in the pupil size) and small eye movements, which may occur during a dynamic sequence acquisition, than the previously proposed textural analysis techniques. The previous gray level co-occurrence matrix algorithm does not provide information about the localization and character of the destabilization of the tear film. Using a block processing technique enables a local identification of tear film disruptions, which may be useful for understanding how the instability of tear film affects the visual outcomes depending on its localization or identifying types of break-up patterns for different scenarios[30]. The spatial resolution will depend on the size of the block and on the separation between the reflected rings. By changing the block size, the algorithm can be adapted to different commercial instruments with variable ring separation.

There have been several works examining automated Tear Film Surface Quality estimators provided in commercially available equipment.[18,19,41,42] However, most of them highly underestimate the expected values of non-invasive break-up time. For example, Yamaguchi et al.[42] reported, for the normal eyes, non-invasive break-up time times between 0 and 6 seconds, for a similar group of normal subjects Hong et al.,[41] found an average non-invasive break-up time of $4.3 \pm 0.3$ seconds. We have reported higher non-invasive break-up time median values for a normal population (15.7 s., 18.43 s. and 16.23 s. for manual, gray level co-occurrence matrix and fractal dimension methods respectively), this values are in accordance with other studies that were using similar technologies[19] and agree with the fact that non-invasive break-up time is higher than fluorescein break-up time inasmuch that fluorescein instillation destabilizes the tear film provoking lower break-up times.[43] One possible reason for this difference in break-up times between studies could be due to the demographics of the studied population or the technical characteristics of the devices used (e.g. source, coverage, software, etc.).

The examined automated methods (fractal dimension and gray level co-occurrence matrix based Tear Film Surface Quality) have shown an excellent and significant correlation with the manually operated evaluation of the recorded sequences $r^2 = 0.86$ ($p < 0.001$) for gray level co-occurrence matrix and $r^2 = 0.82$ ($p < 0.001$) for fractal dimension. Although the correlation with gray level co-occurrence matrix approach is slightly greater, the estimated median difference between the manual and the fractal dimension approach was found to be zero. So in terms of agreement fractal dimension method seems to be somewhat closer to the subjective evaluation. This is supported with the results of Friedman's test where a statistically significant difference was found in non-invasive break-up time measurements between manual and gray level co-occurrence matrix methods but there was no statistically significant difference in non-invasive break-up time measurements between manual and the fractal dimension method.

In this work, the object being measured is the HSV recording while the measuring technique is either a human observer who is assessing it or the proposed automated method for analyzing it. In that sense, the proposed automated method is perfectly repeatable unlike in the case of a human operator that could arrive at different answers every time he/she is presented with the same HSV recording.

One possible weakness of the proposed method which needs to be examined is in relation to its performance on highly irregular corneas, as once encountered in highly astigmatic or keratoconic eyes. In a normal cornea the reflected rings approach a set of concentric circles, therefore when the image is transformed from Cartesian to polar coordinates we obtain a set of quasi-straight horizontal lines. In a highly astigmatic cornea, the reflection of the Placido disk pattern deviates from concentric rings and would need to be appropriately transformed to a new coordinate system, so that the pattern changes related to the corneal irregularity are compensated.[44]

Finally it is worth mentioning that the proposed method of evaluating high speed videokeratoscopy images can be used for subjects with compromised tear film such as those diagnosed with dry eye. In this work, however, only subjects with healthy tear film were considered to establish, in the case of dry eye detection scheme, the performance of the method under the null hypothesis.

Summarizing, assessing tear film stability is essential in dry eye diagnosis. Non-invasive techniques have been proposed as the preferred method for tear film break-up measurements,[6] and in the last years some commercial instruments that incorporate a tear film analysis function have been developed. However, there is a lack of correlation between the proposed automated methods and the manual assessment of tear film stability.

A novel method that allows an automated assessment of tear film stability, which highly agrees with human assessment of high speed videokeratoscopy images, has been developed and evaluated. The presented method is more advantageous than the manual assessment of high speed videokeratoscopy images, since examining manually those images is time consuming, laborious, prone to human error due to fatigue and it cannot be done in real time. By implementing the proposed algorithm in videokeratoscopes, the tear film analysis could be done accurately in real time.

**Acknowledgments**

This project has received funding from the European Union's Horizon 2020 research and innovation program under the Marie Skłodowska-Curie grant agreement No. 642760.

Authors thank Dr. Dorota H. Szczesna-Iskander and Maryam Mousavi for performing clinical assessment of patients and high speed videokeratoscopy images and Dr. David Alonso-Caneiro for providing the custom written graphic user interface used for manual analysis of high speed videokeratoscopy recorded images.